\documentclass[aps,prl,epsfig,showpacs,floats,twocolumn,superscriptaddress,amssymb,amsmath,floatfix]{revtex4}
\usepackage{graphicx}
\begin{document}

%%%%%%%%%%%%%%%%%%%%%%%%%%%%%%%%%%%%%%%%%%%%%%%%%%%%%%%%%%%%%%%%%%%%%%%%
% Macro.TeX  Ver.1
%%%%%%%%%%%%%%%%%%%%%%%%%%%%%%%%%%%%%%%%%%%%%%%%%%%%%%%%%%%%%%%%%%%%%%%%
%
\def\f#1#2{\frac{#1}{#2}}
\def\der#1#2{\f{\D #1}{\D #2}}
\def\fder#1#2{\f{\delta #1}{\delta #2}}
\def\D{\partial}
\def\grad{\nabla}
\def\div{\nabla \cdot}
\def\rot{\nabla \times}
\def\lap{\nabla^2}
\def\inv{^{-1}}
\def\Tr{\mbox{\rm Tr}\;}
\def\const{\mbox{const}}
\def\etal{{\it et al.}}
\def\Av#1{\overline{#1}}
\def\Ref#1{(\ref{#1})}
\def\Eq#1{Eq.(\ref{#1})}
\def\Eqs#1{Eqs.(\ref{#1})}
\def\Ref#1{(\ref{#1})}
\def\Fig#1{Fig.\ref{#1}}
\def\dis{\displaystyle}
\def\eq{\begin{eqnarray}}
\def\qe{\end{eqnarray}}
\def\eqnn{\begin{eqnarray*}}
\def\qenn{\end{eqnarray*}}
\def\nn{\nonumber}
\def\bA{\bm{A}}
\def\bB{\bm{B}}
\def\bC{\bm{C}}
\def\bD{\bm{D}}
\def\bE{\bm{E}}
\def\bF{\bm{F}}
\def\bG{\bm{G}}
\def\bH{\bm{H}}
\def\bI{\bm{I}}
\def\bJ{\bm{J}}
\def\bK{\bm{K}}
\def\bL{\bm{L}}
\def\bM{\bm{M}}
\def\bN{\bm{N}}
\def\bO{\bm{O}}
\def\bP{\bm{P}}
\def\bQ{\bm{Q}}
\def\bR{\bm{R}}
\def\bS{\bm{S}}
\def\bT{\bm{T}}
\def\bU{\bm{U}}
\def\bV{\bm{V}}
\def\bW{\bm{W}}
\def\bX{\bm{X}}
\def\bY{\bm{Y}}
\def\bZ{\bm{Z}}
\def\ba{\bm{a}}
\def\bb{\bm{b}}
\def\bc{\bm{c}}
\def\bd{\bm{d}}
\def\be{\bm{e}}
\def\beff{\bm{f}}
\def\bg{\bm{g}}
\def\bh{\bm{h}}
\def\bi{\bm{i}}
\def\bj{\bm{j}}
\def\bk{\bm{k}}
\def\bl{\bm{l}}
\def\bem{\bm{m}}
\def\bn{\bm{n}}
\def\bo{\bm{o}}
\def\bp{\bm{p}}
\def\bq{\bm{q}}
\def\br{\bm{r}}
\def\bs{\bm{s}}
\def\bt{\bm{t}}
\def\bu{\bm{u}}
\def\bv{\bm{v}}
\def\bx{\bm{x}}
\def\by{\bm{y}}
\def\bz{\bm{z}}
\def\bA{\bm{A}}
\def\bB{\bm{B}}
\def\bC{\bm{C}}
\def\bD{\bm{D}}
\def\bE{\bm{E}}
\def\bF{\bm{F}}
\def\bH{\bm{H}}
\def\bI{\bm{I}}
\def\bJ{\bm{J}}
\def\bK{\bm{K}}
\def\bL{\bm{L}}
\def\bM{\bm{M}}
\def\bN{\bm{N}}
\def\bO{\bm{O}}
\def\bP{\bm{P}}
\def\bQ{\bm{Q}}
\def\bR{\bm{R}}
\def\bS{\bm{S}}
\def\bT{\bm{T}}
\def\bU{\bm{U}}
\def\bV{\bm{V}}
\def\bW{\bm{W}}
\def\bX{\bm{X}}
\def\bY{\bm{Y}}
\def\bZ{\bm{Z}}
\def\ten{$\bullet$ \,}
\def\nten{\noindent $\bullet$ \,}
\def\vsp#1{\vspace{#1 mm}}
\def\hsp#1{\hspace{#1 mm}}
\def\simge{\;\lower3pt\hbox{$\stackrel{\textstyle >}{\sim}$}\;}
\def\simle{\;\lower3pt\hbox{$\stackrel{\textstyle <}{\sim}$}\;}
\def\bm#1{\mbox{\boldmath $#1$}}
\def\tensor#1{{\sf #1}}
\def\lrL#1{\left[#1\right]}
\def\lrM#1{\left\{#1\right\}}
\def\lrS#1{\left(#1\right)}
\def\lrF#1{\left|#1\right|}
\def\lrA#1{\left\langle #1 \right\rangle}
\def\lrAA#1{\left\langle\!\left\langle #1 \right\rangle\!\right\rangle}
\def\biglrL#1{\bigl[ #1 \bigr]}
\def\biglrM#1{\bigl\{#1 \bigr\}}
\def\biglrS#1{\bigl( #1 \bigr)}
\def\biglrF#1{\bigl|#1\bigr|}
\def\biglrA#1{\bigl\langle #1 \bigr\rangle}
\def\bigglrL#1{\biggl[ #1 \biggr]}
\def\bigglrM#1{\biggl\{#1 \biggr\}}
\def\bigglrS#1{\biggl( #1 \biggr)}
\def\bigglrF#1{\biggl|#1\biggr|}
\def\bigglrA#1{\biggl\langle #1 \biggr\rangle}
\def\BigglrL#1{\Biggl[ #1 \Biggr]}
\def\BigglrM#1{\Biggl\{#1 \Biggr\}}
\def\BigglrS#1{\Biggl( #1 \Biggr)}
\def\BigglrF#1{\Biggl|#1\Biggr|}
\def\BigglrA#1{\Biggl\langle #1 \Biggr\rangle}
\def\commentoff#1{}
\def\commenton#1{{\sf #1}}
%
%%%%%%%%%%%%%%%%%%%%%%%%%%%%%%%%%%%%%%%%%%%%%%%%%%%%%%%%%%%%%%%%%%%%
%   Here are the new commands for this document only
%%%%%%%%%%%%%%%%%%%%%%%%%%%%%%%%%%%%%%%%%%%%%%%%%%%%%%%%%%%%%%%%%%%%
\renewcommand{\br}{{\bf r}}
\renewcommand{\bR}{{\bf R}}
\renewcommand{\bx}{{\bf x}}
\renewcommand{\bv}{{\bf v}}
\renewcommand{\bG}{{\bf G}}
\renewcommand{\bH}{{\bf H}}

\newcommand{\tomega}{\Omega}
%%%%%%%%%%%%%%%%%%%%%%%%%%%%%%%%5
%\newcommand{\ord}{\psi}
\newcommand{\bpsi}{\bm{\psi}}
\newcommand{\eRt}{\widetilde{e}_R}
\newcommand{\eRRt}{\widetilde{e}_{RR}}
\newcommand{\eIIt}{\widetilde{e}_{II}}

%%%%%%%%%%%%%%%%%%%%%%%%%%%%%%%%%%%%%%%%%%%%%%%%%%%%%%%%%%%%%%%%%%%%

\title{Many-Body Theory of Synchronization by Long-Range Interactions}

\author{Nariya Uchida}%\email{uchida@cmpt.phys.tohoku.ac.jp}
\affiliation{Department of Physics, Tohoku University, Sendai, 980-8578, Japan}

\pacs{
05.45.Xt,% Synchronization; coupled oscillators
05.40.-a %Fluctuation phenomena, random processes, noise, and Brownian motion 
}

\date{\today}

\begin{abstract}
Synchronization of coupled oscillators on 
a $d$-dimensional lattice with the power-law coupling 
$G(r) = g_0/r^\alpha$ and randomly distributed 
intrinsic frequency is analyzed.
A systematic perturbation theory is developed to 
calculate the order parameter profile and correlation functions
in powers of $\epsilon = \alpha/d-1$. 
For $\alpha \le d$, the system exhibits a sharp synchronization transition 
as described by the conventional mean-field theory.
For $\alpha > d$, the transition is smeared by the quenched disorder, 
and the macroscopic order parameter $\Av\psi$ 
decays slowly with $g_0$ as $|\Av\psi| \propto g_0^2$. 
\end{abstract}

\maketitle

\paragraph{Introduction.}

Collective oscillations of active interacting elements
are observed in a variety of
physical, chemical, and biological systems far from equilibrium.
Numerous studies have been devoted to the mutual entrainment of
oscillators that have different intrinsic 
frequencies~\cite{Kuramoto,Strogatz00,Acebron}.
A class of models with global (or mean-field) coupling 
have enjoyed deep theoretical understanding~\cite{Kuramoto75,Crawford99}.
The phase of the oscillators become coherent 
as the coupling strength $g_0$ exceeds a threshold,
which is the onset of synchronization.
Extensive research has been focused on the 
transition behavior~\cite{Kuramoto75,Crawford99,
Daido94,Crawford95,Hong07}. 
The amplitude of the macroscopic order parameter 
scales as $|\Av\psi| \propto (g_0 - g_c)^\beta$,
with $\beta=1/2$ for the original mean-field model 
by Kuramoto~\cite{Kuramoto75}, while $\beta = 1$ for 
some other types of coupling~\cite{Crawford99,Daido94,Crawford95}. 

Compared to the case of global coupling,
behaviors of locally~\cite{Sakaguchi87,StrogatzMirollo88} 
and non-locally~\cite{Kuramoto95,Strogatz04} coupled oscillators
are still widely open questions. 
In particular, knowledge about synchronization caused by
long-range interactions is quite limited~\cite{Rogers,Marodi,Zaslavsky},
although they are ubiquitous in Nature in the form of,
e.g., gravitational, electromagnetic, elastic, and hydrodynamic forces.
Early numerical works for the power-law coupling $\propto 1/r^\alpha$
in $d$-dimensional array of oscillators show that 
global synchronization 
is possible for $\alpha \le 2$ ($d=1$)~\cite{Rogers}, 
while system-size effect is significant 
for $\alpha \le d$ ($d=1,2$)~\cite{Marodi}.

Recently, we proposed a simple model 
of microfluidic carpets~\cite{hsync-paper1,hsync-paper2}, 
which is
a two-dimensional array of rotors with 
a hydrodynamic coupling $\propto 1/r^3$.
The model exhibits an unconventionally smooth transition 
to the synchronized state~\cite{hsync-paper2}.
The macroscopic order parameter decays gradually 
as the randomness is increased,
in contrast to the sharp transition for global coupling.

Motivated by the numerical results, 
this Letter theoretically addresses synchronization of oscillators 
with a general class of long-range coupling.
We will develop a systematic perturbation
expansion around the mean field,
taking the moments $\sigma_n$ of the interaction $G(\br)$ 
as the small parameters (which is analogous in spirit to the 
cluster expansion in the classical gas theory).
For the power-law coupling $G(\br) = g_0/r^\alpha$,
it is equivalent to a series expansion in $\epsilon = \alpha/d - 1$.
We will solve for the order parameter profile
and correlation functions up to $O(\epsilon^2)$.
The main finding of this paper will be that the macroscopic order parameter
for $\alpha > d$ behaves as $|\Av\psi| \propto g_0^2$ for $g_0 \to 0$,
which means that synchronization persists 
for arbitrary weak coupling.
We interpret it as the result of quenched spatial heterogeneity.
In contrast, for $\alpha \le d$, the heterogeneity is averaged out and
the transition is exactly described by the mean-field theory.

\paragraph{Model.}
In our model,
oscillators indexed by $i = 1,2,\ldots,N$
are arrayed on a $d$-dimensional 
regular lattice with the unit grid size.
The phase $\phi_i$ of the $i$-th oscillator located at $\br_i$ 
obeys the dynamic equation,
\eq
\f{d\phi_i}{dt} &=& \omega_i
- \sum_{j\neq i} G(\br_i - \br_j) 
\sin \left(\phi_i - \phi_j \right),
\qquad
\label{dotphi1}
\qe
where $\omega_i$ is the intrinsic frequency that has
the Gaussian distribution with the standard deviation $\omega_0$,
\eqnn
P(\omega_i) = \f{1}{\sqrt{2\pi}\omega_0} 
\exp\left(-\f{\omega_i^2}{2\omega_0^2}\right).
\qenn
We require the coupling function $G(\br)$ 
to be positive,
%-valued, 
slowly decreasing function of $|\br|$,
so that its moments 
\eqnn
\sigma_n = \sum_{j\neq i} G(\br_i - \br_j)^n
%= \sum_{j \neq i} \frac{g_0^n}{|\br_i - \br_j|^{n\alpha}}.
\qenn
rapidly decays with $n$.
To be specific, let us consider the power-law coupling 
$G(\br) = g_0/r^\alpha$
with the constants $g_0 > 0$ and $\alpha\ge 0$.
We normalize the coupling by rescaling time 
so that $\sigma_1 = 1$ without losing generality. 
%Thus $g_0 = 1/\sum_j |r_i - r_j|^{-\alpha}$ is 
%determined by the lattice and the exponent $\alpha$.
For the global coupling  ($\alpha=0$), 
we have $G(r) = g_0 = 1/N$, and 
the moments $\sigma_n = 1/N^{n-1}$ for $n \ge 2$
vanish as $N\to\infty$.
In more general, for $\alpha < d$, the integral $\int d^d r /r^\alpha$ 
diverges with the system dimension $r_{\rm max} \sim N^{1/d}$,
which means that $g_0 \sim N^{\alpha/d - 1}$
and $\sigma_n$ $(n\ge 2)$ vanish as $N\to \infty$.
This is true also for $\alpha=d$, except that the divergence of $g_0$ 
is logarithmic.
On the other hand, for $\alpha > d$, 
we have $g_0 < 1$ and 
$\sigma_n \approx g_0^n \to 0$ ($n \to \infty$).
Regarding $\epsilon=\alpha/d - 1$ as the small parameter, 
we can show that $\sigma_n = O(\epsilon^n)$.
For example, for $d=1$,
we have 
$g_0 = {1}/{2\zeta(1 + \epsilon)} \approx \epsilon/2\gamma$,
and
$\sigma_n = 2 \zeta(n \alpha) g_0^n \approx 
2 \zeta(n) (\epsilon/2 \gamma)^n $ for $n\ge 2$,
where $\gamma$ is Euler's constant.
Our perturbation theory will be given as 
a series expansion in $\epsilon$ via $\sigma_n = O(\epsilon^n)$.

\paragraph{Order Parameter.}

In order to describe the collective behavior,
%synchronization transition,
we introduce the site-dependent complex order parameter $\psi_i$
with its amplitude $\rho_i$ and phase $\theta_i$~\cite{Kuramoto95}, 
\eq
\psi_i = \rho_i e^{i\theta_i}= \sum_{j\neq i} G(\br_i-\br_j) e^{i\phi_j},
\label{RandTheta}
\qe
with which we can rewrite (\ref{dotphi1}) as
\eq
\der{\phi_i}{t} &=& \omega_i - \rho_i \sin (\phi_i - \theta_i).
\label{dotphi2}
\qe
Note that $\rho_i \le 1$  due to the normalization of $\sigma_1$.
When the coupling is long-ranged, $\psi_i$ 
involves infinitely many 
oscillators and is expected to change much slower than $\phi_i$.
Therefore, we approximate $\psi_i$ to be constant in time.
Then Eq.(\ref{RandTheta}) is replaced by its temporal average,
\eq
\psi_i = 
\rho_i e^{i\theta_i}
= \sum_{j} G_{ij} e^{i\theta_j} E(\rho_j, \omega_j),
\label{Rtav}
\qe
where 
$G_{ij} = G(\br_i - \br_j)$ for $i\neq j$, and $G_{ij} = 0$ for $i=j$.
The function 
$E(\rho_j, \omega_j)$ 
is the temporal average of $e^{i(\phi_j-\theta_j)}$,
and is calculated following the original prescription 
by Kuramoto~\cite{Kuramoto75,Kuramoto}.
First, for an oscillator that satisfies 
$|\omega_i|<\rho_i$ ({\it coherent case}), 
Eq.(\ref{dotphi2}) allows the stationary solution 
$
\phi_i = \theta_i + \sin^{-1} \left( {\omega_i}/{\rho_i}\right)
$,
which gives
\eqnn
E(\rho_j, \omega_j) =
e^{i(\phi_j - \theta_j)} =  
\sqrt{1-\f{\omega_j^2}{\rho_j^2}} + i \f{\omega_j}{\rho_j}.
\label{Ecoh}
\qenn
On the other hand, if $|\omega_i|>\rho_i$ ({\it incoherent case}), 
Eq.(\ref{dotphi2})
has a drifting solution, which visits each value of $\phi_i$ 
with the frequency that is inversely proportional to the angular velocity:
$
\nu(\phi_i) 
= \nu_i |\dot{\phi}_i|^{-1} 
= \nu_i \left|\omega_i - \rho_i \sin(\phi_i - \theta_i)\right|^{-1}
$.
Here, 
% $\nu_i$ is the normalization constant given by
the constant $\nu_i = \f{1}{2\pi} \sqrt{\omega_i^2 - \rho_i^2}$
ensures the normalization $\int_0^{2\pi} d\phi \,\nu(\phi) = 1$.
It gives the temporal average 
$\int_0^{2\pi} d\phi \,\nu(\phi )e^{i(\phi_j - \theta_j)}$ as
\eqnn
E(\rho_j, \omega_j)
&=&  
i \f{\omega_j}{\rho_j} \left(1 - \sqrt{1- \f{\rho_j^2}{\omega_j^2}} \right).
\label{Eincoh}
\qenn

\paragraph{Perturbation Expansion.}

\begin{figure}[t]
\includegraphics[width=0.99\columnwidth]{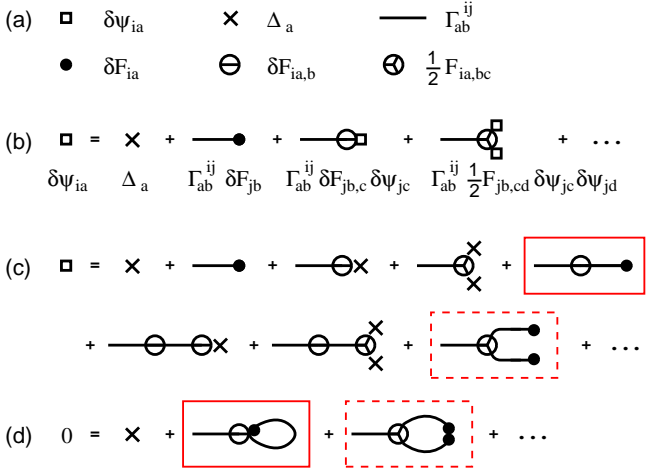}%diagramRule2.eps}
\caption{
Diagrammatic expression of the perturbation scheme.
(a) Definition of symbols. 
(b) Expression of Eqs.(\ref{dPa2},\ref{defT})
and (c) its recursive expansion.
(d) The ensemble average of (c). 
See text for interpretation of the framed graphs.
}
\label{figDiagram}
\end{figure}

Let us introduce the two-dimensional vector 
$\bpsi_i = (\psi_{iR}, \psi_{iI})
= (\rho_i \cos \theta_i, \rho_i \sin \theta_i)$,
to rewrite Eq.(\ref{Rtav}) in the vectorial form
\eq
\bpsi_i &=& \sum_{j} G_{ij} \bm{F}(\bpsi_j, \omega_j),
\label{bord}
\\
\bm{F}(\bpsi_j, \omega_j) &=& 
\left(\begin{array}{c} \mbox{Re} \\ \mbox{Im} \end{array}\right)
e^{i\theta_j} E(\rho_j, \omega_j),
\label{Ford}
\qe
Our task is to calculate the spatial average of the order parameter,
\eqnn
\Av{\psi}_a = \f{1}{N}  \sum_i \psi_{ia},  \quad a = R,I,
\qenn
which is equivalent to the ensemble average over $\omega_i$'s.
Expecting that spatial fluctuation of the order parameter
is small for long-range interactions,
we expand the RHS of (\ref{bord}) 
with respect to the deviation $\delta \bpsi_j = \bpsi_j - \Av{\bpsi}$, as
\eq
\psi_{ia} = 
G_{ij} 
\left[
F_{ja} + F_{ja,b} \, \delta\psi_{jb} + \f{F_{ja,bc} }{2} \, \delta\psi_{jb}\delta\psi_{jc} + \ldots
\right],
\label{Pa}
\qe
with
$F_{ja} = F_a(\Av{\bpsi}, \omega_j)$,
$F_{ja,b} = \der{}{\Av\psi_b} F_{ja}(\Av{\bpsi}, \omega_j)$,
and
$F_{ja,bc} = 
\der{^2}{\Av\psi_b \D \Av\psi_c} F_{ja}(\Av{\bpsi}, \omega_j)$.
Here and hereafter, summation over repeated indices $a,b,c,d = R, I$
and $i,j,k, \ell=1,2, \ldots, N$ are implied.
We decompose the zeroth and first order coefficients
into their averages 
$f_a = f_a(\Av{\bpsi}) = \lrA{F_{ja}}_{\omega_j}$, 
$f_{a,b} = \f{\D f_a}{\D \Av\psi_b}$
and the deviations 
$\delta F_{ja} = F_{ja} - f_a$,
$\delta F_{ja,b} = F_{ja,b} - f_{a,b}$.
Subtracting $\Av{\psi}_a$ from (\ref{Pa})
and then multiplying by the inverse of
the $2N \times 2N$ matrix
$
M^{ij}_{ab} = \delta_{ij} \delta_{ab} - G_{ij} f_{a,b}
$,
which is expanded as
$
\lrL{M^{-1}}^{ij}_{ab} = 
\delta_{ij} \delta_{ab} 
+ G_{ij}     f_{a,b} 
+ G^2_{ij} f^2_{a,b} 
+ \ldots
$
with $G^2_{ij} = G_{ik} G_{kj}$, $f^2_{a,b} = f_{a,c} f_{c,b}$, etc.,
we obtain
\eq
\delta\psi_{ia} 
&=& \Delta_a + \Gamma_{ab}^{ij} U_{jb},
\label{dPa2}
\\
U_{jb} &=&
\delta F_{jb} + \delta F_{jb,c} \delta\psi_{jc}
+ \f12 F_{jb,cd} \delta\psi_{jc}\delta\psi_{jd}
+ \ldots,
\label{defT}
\\
\Delta_a &=& \left(I_2 - \textstyle \der{f}{\Av\psi}\right)^{-1}_{ab} 
\left(f_b - \Av\psi_b \right),
\label{defDelta}
\\
\Gamma_{ab}^{ij} &=& 
\lrL{ M^{-1} (G \otimes I_2)}_{ab}^{ij} =
G_{ij} \delta_{ab} + G^2_{ij} f_{a,b} 
+\ldots,
\qquad
\label{defGamma}
\qe
where we used the $2\times 2$ matrices
$I_2 = \{\delta_{ab}\}$ and $\der{f}{\Av\psi} = \{ f_{a,b}\}$.
Eqs.(\ref{dPa2},\ref{defT}) 
can be diagrammatized as shown in Fig.\ref{figDiagram}(b),
by combining the symbols defined in Fig.\ref{figDiagram}(a).
Recursively using (\ref{dPa2}) for the $\delta \bpsi$'s in (\ref{defT}),
we get an expansion of $\delta \psi_{ia}$
in terms of $\Delta_a$, $\Gamma_{ab}^{ij}$, $\delta F_{ja}(\Av\bpsi)$, $F_{ja,bc}(\Av\bpsi)$
and their derivatives; see Fig.\ref{figDiagram}(c).
The terminators $\Delta_a$ and $\delta F_{ja}$
are connected by the vertices $\delta F_{ja,b}, F_{ja,bc}, \ldots$ 
and propagators $\Gamma_{ab}^{ij}$ to the site~$i$.
For example, the graph framed by solid lines reads
$\Gamma_{ab}^{ij} \cdot \delta F_{jb,c} 
\cdot \Gamma_{cd}^{jk} \delta F_{kd}$,
and the dot-framed graph reads
$\Gamma_{ab}^{ij} \cdot \f12 F_{jb,cd} 
\cdot \Gamma_{ce}^{jk}    \delta F_{ke} 
\cdot \Gamma_{df}^{j\ell} \delta F_{\ell f}$.

Now we take the average of \Eq{dPa2} over the distribution of $\omega_i$'s.
The LHS vanishes by the definition of $\delta \bpsi$.
On the RHS, 
$\delta F_{jb}$ and its derivative
$\delta F_{jb,c}$ are averaged out
unless they are correlated with a partner at the same site. 
Graphically, it means that the legs of the graphs 
(with the black dots at their ends) have to be attached 
to each other or to vertices 
to produce correlation terms.
For example, 
%the graph in the solid frame in Fig.\ref{figDiagram}(c) is 
%averaged out because of the factor $\lrA{\delta F_{jb,c}}=0$.
%On the other hand, 
the dot-framed graph in Fig.\ref{figDiagram}(c) yields
the corresponding graph in Fig.\ref{figDiagram}(d), which reads
$
\Gamma_{ab}^{ij} 
\cdot \f12 \lrA{F_{jb,cd} }_j \cdot
\Gamma_{ce}^{jk}    \Gamma_{df}^{jk} 
\lrA{ \delta F_{ke} \delta F_{k f}}_k 
$,
where $\lrA{\cdots}_j$ means the average over the distribution $P(\omega_j)$.
%Note that the $\delta F_{ke}$ and $\delta F_{\ell f}$ in the 
%original graph are now paired at $k=\ell$.
%Expanding $\Gamma$'s in powers of $G_{ij}$ using (\ref{defGamma}), 
Using the expansion (\ref{defGamma}) with the trace $G^n_{jj} = \sigma_n$, 
we obtain the $O(\epsilon^2)$ expression of this graph as
\eqnn
\f{\sigma_2}{2} f_{a,cd} \left(g_{cd} - f_c f_d\right),
\label{AvDotted}
\qenn
where the functions
$f_{a,cd}(\Av\bpsi) = \der{^2 f_a}{\psi_c \D \psi_d}$
and 
$g_{ab}(\Av\bpsi) = \lrA{F_{ka} F_{kb}}_k$
are introduced.
Another graph that gives an $O(\epsilon^2)$ contribution is
framed by solid lines in Fig.\ref{figDiagram}(d). It reads
$\Gamma_{ab}^{ij} \Gamma_{cd}^{jj} \lrA{ \delta F_{jb,c} \delta F_{jd}}_j
$ 
and is approximated as
\eqnn
\sigma_2 f_{c,d} \left( h_{da,c} - f_d f_{a,c} \right)
\label{AvSolid}
\qenn
with the function
$h_{ab,c}(\Av\bpsi) =  \lrA{F_{ja} F_{jb,c}}_j$.
%Here we used the expansion of the self-looped
%propagator $\Gamma_{cd}^{jj}
%= \sigma_2 f_{c,d} + \sigma_3 f^2_{c,d} + \ldots$.
We can see that these two graphs and $\Delta_a$ 
are the only $O(\epsilon^2)$ contributions.
Combining them and using Eq.(\ref{defDelta}), 
we obtain the self-consistent equation for $\Av\bpsi$ 
to $O(\epsilon^2)$ as
\eq
\Av{\psi}_a &=&
f_a + \sigma_2 (\delta_{ab} - f_{a,b}) V_b,
\label{sceqP}
\\
V_b
&=&
f_{c,d} \left( h_{db,c} - f_d f_{b,c} \right) 
+ 
\f12 f_{b,cd} \left(g_{cd} - f_c f_d\right).
\label{sceqQ}
\qe

\paragraph{Correlation Function.}

The correlation function of the order parameter
$C_{ab}^{ij} = C_{ab}(\br_i - \br_j) 
= \lrA{\delta \psi_{ia} \delta \psi_{jb}}
$ 
can be also computed using the diagrams.
There is only one non-vanishing graph at $O(\epsilon^2)$, 
which gives
\eq
C_{ab}^{ij} 
%&=& \Gamma_{ik}^{ac} \Gamma_{kj}^{cb} 
%\lrA{\delta F_{kc} \delta F_{kc}}_k
\simeq
G^2_{ij} \lrS{g_{ab} - f_a f_b}. 
\label{Cij}
\qe
Note that $G^2_{ij} = G_{ik} G_{kj}$
is a function of $\br_{ij} = \br_i - \br_j$. 
At large distance, it decays 
as $G^2_{ij} \propto |\br_{ij}|^{-d(1+ 2\epsilon)}$ for $\epsilon>0$,  
as we can see from a simple dimensional analysis.
(For $d\ge 2$, $G^2_{ij}$ depends also on the direction of $\br_{ij}$
reflecting the lattice anisotropy).
On the other hand, setting $i = j$ in (\ref{Cij}), 
we obtain the variance of the order parameter,
\eq
\lrA{\delta\psi_{ia}\delta\psi_{ib}} = \sigma_2 \lrS{g_{ab} - f_a f_b}.
\label{varP}
\qe

\paragraph{Transition Behavior.}

\begin{figure}[h]
\includegraphics[width=0.8\columnwidth]{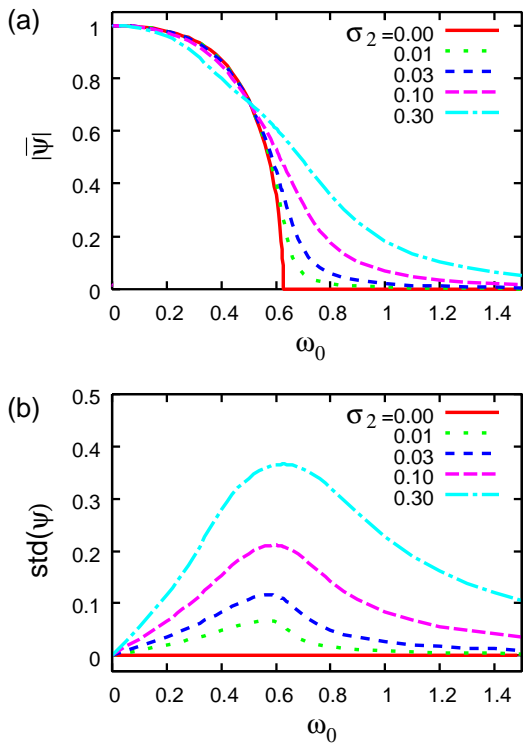}%fig2V.eps}
\caption{
(a) The macroscopic order parameter $|\Av{\psi}|$ and
(b) the standard deviation std$(\psi) = \lrA{|\delta\psi|^2}^{1/2}$
as functions of $\omega_0$. 
The long tails scale as $|\Av\psi| \propto \sigma_2/\omega_0^2$
and std$(\psi) \propto \sigma_2/\omega_0$.
}
\label{figAvP}
\end{figure}

In order to solve the self-consistent equation (\ref{sceqP},\ref{sceqQ}),
we need to compute
$f_a$, $g_{ab}$, $h_{ab,c}$ and their derivatives 
as functions of $\Av\psi = \Av\rho e^{i\Av\theta}$.
To simplify calculations,
we choose the coordinate frame in which $\Av{\theta} = 0$.
%Averaging Eq.(\ref{Ford}) over the distribution of $\omega_j$, we have 
Then the ensemble average of Eq.(\ref{Ford}) gives
\eq
f_a &=&  e_a(\Av\rho)
%,\qquad
%e_a(\Av\rho) 
= \int_{-\infty}^\infty d\omega P(\omega) E_a(\Av\rho,\omega)
\label{ea}
\qe
$(a=R,I)$.
Here, $E_R(\rho,\omega)$ and $E_I(\rho,\omega)$ 
are the real and imaginary parts of $E(\rho, \omega)$, respectively.
Note that $f_I = e_I(\Av\rho)=0$ thanks to
the parity of $P(\omega)$ (even) and $E_I(\rho,\omega)$ (odd).
The quadratic moments read
\eq
g_{RR} &=& e_{RR}(\Av\rho), 
\quad
g_{RI} = g_{IR} = 0,
\quad
g_{II} = e_{II}(\Av\rho),
\nn\\
e_{ab}(\Av\rho) &=& \int_{-\infty}^\infty d\omega P(\omega) 
E_a(\Av\rho,\omega) E_b(\Av\rho,\omega).
\label{eab}
\qe
The calculations of the derivatives $f_{a,b}$, $f_{a,bc}$ and 
$h_{ab,c}$ are also straightforward.
% and will be published elsewhere.
The non-vanishing components are found to be 
$
f_{R,R} = e'_R, \, 
f_{I,I} = \eRt, \,
f_{R,RR} = e''_R, \,
f_{R,II} = f_{I,RI} = f_{I,IR} = \eRt', \,
$
and 
$
h_{RR,R} = e'_{RR}/{2}, \,
h_{RI,I} = \eRRt, \, 
h_{IR,I} = -\eIIt,\,
h_{II,R} = e'_{II}/2, \,
$
where $'= d/d\rho$ and the abbreviations 
$\eRt =  e_R/\Av\rho$,  
%$\eRt' = e'_R/\Av\rho - e_R/{\Av\rho^2}$,
$\widetilde{e}_{RR} = e_{RR}/\Av\rho$, 
and 
$\widetilde{e}_{II} = e_{II}/\Av\rho$
are used.
%All the functions are evaluated at $\rho = \Av\rho$.
Substituting these into Eqs.(\ref{sceqP},\ref{sceqQ}),
we obtain
\eq
\Av{\psi}_R &=& 
e_R + \sigma_2 (1 - e'_R) V_R,
\label{sceqPR}
\qquad
\\
V_R &=&
\f12 \Big[
e'_R \left(e'_{RR} - 2 e_R e'_R \right)
- 2 \eRt \widetilde{e}_{II}
\nn\\&& \quad\qquad 
+ \, e''_R \left( e_{RR} - e_R^2 \right) + \eRt' e_{II}
\Big]. 
\label{sceqQR}
\qe
and $\Av{\psi}_I = 0$.
On the RHS of (\ref{sceqPR}) are functions of $\Av{\rho}$,
which is related to $\Av\psi_R = \Av{\rho\cos\theta}$ on the LHS
via the expansion 
%\eq
$
\Av{\rho} = \lrA{\sqrt{\psi_R^2 + \psi_I^2}}
= \Av{\psi}_R  + 
{\lrA{\delta \psi_I^2}}/{\Av{\psi}_R} 
+ O(\epsilon^3).
\label{RhoinPR}
$
%\qe
Using this in the RHS of (\ref{sceqPR})
%and borrowing 
with the result $\lrA{\delta \psi_I^2} =\sigma_2 e_{II}$ 
taken from Eq.(\ref{varP}), we arrive at 
the final form of the $O(\epsilon^2)$ self-consistent equation,
\eq
\Av\psi_R &=& e_R + \sigma_2 
\left[
(1- e'_R) V_R + e'_R \widetilde{e}_{II}
\right],
\label{sceqPRfinal}
\qe
with the functions on the RHS 
evaluated at $\rho = \Av\psi_R$.
Its solution gives the order parameter profile
$\Av\psi_R=\Av\psi_R(\omega_0)$.
For $\sigma_2=0$, or $\alpha \le d$, 
Eq.(\ref{sceqPRfinal}) reduces to the mean-field equation 
$\Av\psi_R = e_R(\Av\psi_R)$~\cite{Kuramoto,Kuramoto75}.
The Taylor expansion 
$e_R(\rho) \approx (\omega_c/\omega_0) (\rho - \rho^3/8)$
with $\omega_c = \sqrt{{\pi}/{8}} \simeq 0.627$
reproduces the global-coupling result 
that the order parameter 
vanishes for $\omega_0 > \omega_c$.
In contrast, for $\sigma_2 > 0$, or $\alpha > d$,
there is no sharp transition, and 
the order parameter exhibits a long tail at large $\omega_0$,
In fact, the approximation $e_{RR}(\rho) \approx 
e_{II}(\rho) \approx (8\omega_c/3\pi \omega_0) \rho$
for $\rho\ll 1$ gives the asymptotic behavior of 
$\Av\psi_R$ for $\omega_0 \gg \omega_c$,
\eq
\Av\psi_R \approx 
\f{\sigma_2}{6 \omega_0^2}.
%\left(1+ \frac{\omega_c}{\omega_0} \right).
\label{asympPR}
\qe
The complete order parameter profile is 
obtained by numerical computation of
the functions $e_a(\rho)$ and $e_{ab}(\rho)$, 
and is shown in Fig.\ref{figAvP}(a).
Note that $\Av\psi_R$ in the current coordinate frame
corresponds to $|\Av\psi|$ in the general frame.
As we can see, the deviation from the mean-field profile is 
significant even for relatively small values of $\sigma_2$.
(For comparison, $\sigma_2 = 0.2$ for $(d, \alpha)=(1, 2)$ 
and $\sigma_2 \simeq 0.057$ for $(d,\alpha)= (2, 3)$ (square lattice).) 
The macroscopic order parameter is larger than
the mean-field value for $\omega_0 > \omega_t \simeq 0.504$,
and smaller for $\omega_0 < \omega_t$ for any non-zero value of $\sigma_2$.
The enhancement of synchronization for large $\omega_0$
might look counter-intuitive, but it is a
natural result of the spatial heterogeneity;
there are regions that are more uniform than the others
in terms of the intrinsic frequencies of the oscillators they contain.
These regions can remain synchronized when the other regions are
desynchronized, and contribute to the long tail of  
the order parameter profile.
This effect of quenched heterogeneity
is averaged out in the global-coupling case.
Note also that we have rescaled the timescale 
%and the coupling strength
so that $\sigma_1 = 1$. 
If $\sigma_1$ is not normalized, 
we must divide the intrinsic frequency
and the order parameter by $\sigma_1$,
which modifies Eq.(\ref{asympPR}) as
$|\Av\psi| \approx {\sigma_1 \sigma_2}/{6\omega_0^2} 
= C g_0^2/\omega_0^2$, where 
$C$ is a function of $\alpha$ and $d$.

%

% I would like to briefly mention that
It should be briefly mentioned that
the self-consistent solution bifurcates 
at very small $\omega_0$, into
two stable branches $\Av\psi_{R1} \simeq 1$ and $\Av\psi_{R2} \ll 1$.
The threshold $\omega_b$ rises with $\sigma_2$;
e.g., $\omega_b = 0.003$ for $\sigma_2 = 0.1$
and $\omega_b = 0.022$ for $\sigma_2 = 0.3$.
However, it turns out that the lower branch 
does not satisfy the condition for 
the series (\ref{defGamma}) to converge.
It converges with its trace $\sum_{n=0}^\infty \sigma_n f^n_{a,a}$ 
if $\mbox{max} (e'_R(\Av\psi_R), \tilde{e}_R(\Av\psi_R)) < 1/g_0$.
Plotted in Fig.\ref{figAvP}(a) is 
the upper branch, which always meets the condition.

The standard deviation std$(\psi) = \lrA{|\delta\psi|^2}^{1/2}$
is readily calculated from Eqs.(\ref{varP},\ref{ea},\ref{eab}), 
and is plotted in Fig.\ref{figAvP}(b).
For any non-zero value of $\sigma_2$,
it exhibits a peak near $\omega_0=\omega_c$ 
and a long-tail for $\omega_0 \gg \omega_c$.
The asymptotic behavior for large $\omega_0$ is obtained
via the Taylor expansion of $e_R(\rho)$, $e_{RR}(\rho)$ 
and $e_{II}(\rho)$, as
std$(\psi) \approx \sigma_2/3\sqrt{\omega_c}\omega_0$.

\paragraph{Summary.}
We have found that the mean-field picture of sharp synchronization 
transition is valid only for $\alpha \le d$, 
and the transition is broadened for $\alpha > d$.
It could be regarded as a novel example of smeared transition in 
random systems, which usually requires 
spatially correlated disorder~\cite{Vojta}.
The limitations of the perturbation theory for large $\alpha$
should be assessed by analysis of higher order corrections
and comparison with numerical results, which are beyond 
the scope of the present paper and will be discussed elsewhere.

\acknowledgments

I wish to thank Ramin Golestanian for useful comments, discussions, 
and collaborated works that motivated the present study.

\end{document}